\documentclass{article}
\usepackage[utf8]{inputenc}
\usepackage{natbib}
\usepackage{amsmath}
\usepackage{amssymb}
\usepackage{listings}
\usepackage{graphicx}
\usepackage{amssymb}
\usepackage{epstopdf}
\usepackage{color}
\usepackage{float}
\usepackage{tikz}
\usepackage{tcolorbox}
\usepackage{multirow}
\usepackage{hhline}
\usepackage{xcolor}
\usepackage{colortbl}
\usepackage{graphicx}
\usepackage{subcaption}
\usepackage{booktabs}
\usepackage{amsmath}
\usepackage{algorithm}
\usepackage{algpseudocode}
\newcounter{algsubstate}
\renewcommand{\thealgsubstate}{\alph{algsubstate}}

\usepackage[all]{xy}
\usepackage[affil-it]{authblk}
\usepackage[doublespacing]{setspace}
\usepackage{rotating}

\algnewcommand\algorithmicforeach{\textbf{for each}}
\algdef{S}[FOR]{ForEach}[1]{\algorithmicforeach\ #1\ \algorithmicdo}

\usepackage{hyperref}	
\hypersetup{colorlinks,linkcolor={blue},citecolor={blue!50!black},urlcolor={blue}}

\usepackage{geometry}
\usetikzlibrary{arrows, backgrounds, automata, fit, scopes, calc, matrix, positioning, intersections, decorations.pathmorphing, patterns, arrows.meta, decorations.pathreplacing, patterns, shapes, calligraphy, trees}

\title{Individualized conformal}
\author{Fernando Delbianco \ - \ Fernando Tohm\'e \\ 
Departamento de Econom\'{\i}a, Universidad Nacional del Sur \\
Instituto de Matem\'atica de Bah\'{\i}a Blanca, UNS - CONICET}
\date{}
\begin{document}

\maketitle
\begin{abstract}
    The problem of individualized prediction can be addressed using variants of conformal prediction, obtaining the intervals to which the actual values of the variables of interest belong. Here we present a method based on detecting the observations that may be relevant for a given question and then using simulated controls to yield the intervals for the predicted values. This method is shown to be adaptive and able to detect the presence of latent relevant variables.
\end{abstract}

\section{Introduction}

We present a novel approach to {\em individualized conformal prediction}, intended to yield the intervals in which the actual predicted values of certain variables might be found. We proceed by first detecting {\em relevant} observations for a given target {\em query} using a technique reminiscent to {\em divide and conquer} (\cite{chen2021divide}). Then we apply a data augmentation procedure similar to the generation of {\em repro samples} (\cite{xie2022repro}), simulating controls as if they were bootstrapped (as in \cite{tran2017bayesian}). The development of this approach is based on the statistical technique known as {\it Conformal Prediction}.\\

Conformal Prediction (CP) is a distribution-free prediction methodology usually based on machine learning systems. CP generates predictions about new test data points on the basis of training labeled datasets, only assuming the exchangeability of the data. It requires a pre-specified level that restricts the frequency of errors that the algorithm is allowed to make (see, for instance, \cite{vovk2005algorithmic}, \cite{xie2022homeostasis}, \cite{lei2018distribution}, \cite{barber2021predictive}). Variants of this methodology are {\em split} and {\em jacknife} distribution-free conformal inference (\cite{lei2018distribution}). A recent development that is closely related to our contribution is the adaptive version of CP, based on the application of self-supervised learning (\cite{seedat2023improving}). \\

Another strand in the literature that is relevant for our approach is {\em individualized inference}. This is a particularly useful methodology in the case of large datasets, since it addresses the heterogeneity of data. One approach is based on generating {\em iGroups} of data sharing common features with targeted individuals (\cite{cai2021individualized}). Alternatively, in the {\em iFusion} approach the group of relevant data is generated by fusing the inferences from individuals that are similar to the targeted one (\cite{shen2020fusion}). Individualized inference is also invoked to categorize cases by mixing Gaussian processes generated by individuals with related features (\cite{alaa2017bayesian}). \\

The specific techniques used to carry out the inferences are varied. \cite{chernozhukov2017double} develops a Debiased Machine Learning method to obtain inferences about specific parameters. More traditional methods can also be applied, like inferring {\it confidence} (\cite{xie2013confidence}, \cite{schweder2016confidence}) and  {\it predictive}  (\cite{shen2018prediction}) distributions, which could be seen as ``Bayesian-like frequentist techniques'' that derive distributions up from observed data. In practical terms, an important contribution is the implementation in the {\bf R} language of trainable $p$-value {\em functions} (instead of just $p$-values) by \cite{infanger2019p}. \\

Our own approach has the following features: 
\begin{itemize}
\item It is adaptive. 
\item It admits new unlabeled data as input.
\item It satisfies the condition of {\em exchangeability} by restricting the focus on relevant, and then simulated, data. 
\item It detects empirically the presence of {\em latent} relevant variables, by implicitly getting rid of the factor that generates a failure of the i.i.d. condition.
\end{itemize}

This paper is structured as follows. We first present, in Section 2, the motivation of our proposal. Section 3 details a procedure implementing the method developed. Section 4 presents the empirical setting in which we test the methodology and the corresponding results. Section 5 concludes by presenting ideas for further developments.\\

\section{Motivation}

We assume, as in \cite{delbianco2021methodology}, a statistical model of a data generation process, which can be described as $\{\mathcal{O},\mathcal{P}\}$, where $\mathcal{O}$ is the set of observations while $\mathcal{P} = \{ P_{\theta}: \theta \in \Theta \}$ is a family of probability distributions over $\mathcal{O}$ and $\Theta$ is the space of parameters of the model. The goal is again to estimate intervals for the parameters in response to any {\em query} q, where q is a specific request for information under a given inference method $\mathcal{I}$ applied on the database. \\

The query defines several dimensions. First, $\mathcal{O}$ consists of entries $\{o_i = \langle x_i,y_i \rangle\}_{i=1, \ldots, n}$, where $x_i$ is a matrix of $p$ variables and $n$ observations, the {\em tail} of $o_i$, while $y_i$ is the {\em head}, a vector of dimension $n$. The query $q$ consists of a {\em tail}, $x_0$, with no {\em head}, and with $x_o \neq x_i$, for $i = 1, \ldots, n$. We assume that there exists a class of {\em latent variables} $\mathcal{S}$ such that given a query $q$ there exists a corresponding $s_q \in \mathcal{S}$ yields a class of {\em relevant} observations $\mathcal{O}_q \subseteq \mathcal{O}$ \footnote{Another reason to use controls from a subset of the observations is due the notion of the Law of Large Populations.  \cite{meng2018statistical} refers to the difference between data \textit{quantity} and data \textit{quality}.}. \\

\begin{figure}[h!]
		\centering 
	\caption{Latent relevant area $S_q$} 
		\begin{tikzpicture}[scale=1]
						
	\draw[thick, stealth-stealth] (0,5) node[left] {$y$} -- (0,0) -- (6,0) node[below] {$x$};
	
	\draw[thick,dashed] (3.5,3) circle (1cm);
	
	\draw[dotted] (0,2) -- +(3.5,0);
	\draw[dotted] (0,4) -- +(3.5,0);
	
	\draw (0,2) -- +(-3pt,0) -- +(3pt,0);
	\draw (0,4) -- +(-3pt,0) -- +(3pt,0);
	
	\draw (3.5,0) -- +(0,-3pt) node[below] {$q$} -- +(0,3pt);
	
	\draw (0,4) ++(-6pt,0) -- ++(-6pt,0) -- ++(0,-2) -- ++(6pt,0);
	
	\draw[thick, -stealth] (3.5,0.15) -- (3.5,1.9);
		
		\end{tikzpicture}
    \caption*{A particular value of a tail $q$, is associated to a latent variable $s_q$ that yields the relevant space of tails $x \in X_q$ and as a consequence allows to infer the corresponding class of heads $y \in Y_q$. Then, this bi-dimensional figure can be extended to a tri-dimensional one, with the sequence of different queries as a third axis. This represents how the relevant set varies according to $q$, learning the associated $\theta^q$.}
    \label{fig:latent}
\end{figure}

We then generate a class of controls  $\bar{\mathcal{O}}_q$ verifying that $\mathcal{O}_q\subseteq \bar{\mathcal{O}}_q$. Based on $\bar{\mathcal{O}}_q$ the application of the inference procedure $\mathcal{I}$ yields an interval $\mathbf{I}_q$, with $\theta^q \in \mathbf{I}_q$ such that $\bar{\mathcal{O}}_q$ can be understood as a set of draws from a distribution $\bar{P}_{\theta^q} \in \mathcal{P}$ \footnote{Also known as \textit{Transitional inference}, as in \cite{li2021multi}.}. The relevant set of observations is given by $\mathcal{O}_q = f_{\mathcal{S}}(s_q)$, where $f_{\mathcal{S}}:  \mathcal{S} \rightarrow 2^{\mathcal{O}}$ characterizes a {\em selection} procedure.  \\

\begin{figure}[h!]
		\centering 
		\caption{Simulated controls $\in \bar{\mathcal{O}}_q$} 
		\begin{tikzpicture}[scale=1]
				
	\draw[thick] (2.5,0) ellipse (2 and 0.75);
	
	\node[below] at (4.5,-0.25) {$\mathcal{O}$};
	
	\draw[thick] (2.25,0) circle (0.5cm);
	
	\node[right] at (2.75,0) {$\mathcal{O}_q$};
	
	\draw[thick] (2.5,3) ellipse (2 and 0.75);
	
	\draw[dashed] (1.75,0) -- (0.5,3);
	\draw[dashed] (2.75,0) -- (4.5,3);
	
	\node[below] at (4.5,2.75) {$\bar{\mathcal{O}}_q$};
	\node[above] at (0.5,3.25) {};
	
	\draw[thick, -stealth] (4.75,3) to[out=25,in=155] +(2,0) node[right] {$\bar{P}_{\theta^q}$};
	
	\draw[thick, -stealth] (4.75,0) to[out=25,in=155] +(2,0) node[right] {$P_{\theta^q}$};
		\end{tikzpicture}
		\caption*{The relevant set $\mathcal{O}_q$, detected by means of $f_{\mathcal{S}}$, is extended to an enlarged set, $\bar{\mathcal{O}}_q$. The difference in the inference based in new controls can be measured as $\Delta_{\theta^q}=E(\bar{P}_{\theta^q})-E(P_{\theta^q})$.}
        \label{fig:simulated}
\end{figure}

For an example consider the case of different economies $i$, each one described by a vector of a few macroeconomic variables $o_i = \langle x_i, y_i \rangle$, where $y_i$ is the GDP of country $i$. We can ask, for any given economy $\bar{o} \notin \mathcal{O}$ what is its expected GDP in five years, knowing only $\bar{x}$. Now assume that a latent variable is the productivity of the leading sectors in an economy. Then we can group all the countries in $\mathcal{O}$ with a similar productivity as that of $\bar{o}$ and generate a class of controls based on this choice. \\

Taking another example, we can consider a dataset of programs, students and grades. For a new student, we can ask different questions. For instance, how long will take for a new student to finish her studies? What is the probability of her changing majors? or the probability of dropping out school?, among many other possible queries about this particular student that could be answered with the observations in the dataset. Of course, a different query may require different controls. And this can be true for both the variables and the observations. So the $s_q$ latent variable will yield the portion of observations that is relevant given $q$, and will let us estimate $f_{\mathcal{S}}$ in order to simulate new controls and gain robustness in the inference. \\

To proceed in this way we need to specify, for each $q$ a latent variable (or set of latent variables) that are relevant for the query. Then, we have to distinguish its range using some proxy or measurement on the available $\mathcal{O}$. This will yield the class of relevant observations. For the sake of simplicity, we could assume that exists only two types of query, $q_A$ and $q_B$, and the latent variable $s_q$ defines $\mathcal{O}_{q_A}$ and $\mathcal{O}_{q_B}$. This setting can be later generalized to cases of more than two types of query. In the particular example of the students, let us assume that there are two types of students, associated to a latent variable (which can capture, for instance, socioeconomic or cognitive advantages). Then, each $q$ will be associated to its corresponding type of student and according to $f_{\mathcal{S}}$, mapped to the class of observations of students of that type. \\

As said, the class $\mathcal{O}_q$ is that of the entries in $\mathcal{O}$ that are {\em relevant} to answering $q$. But once obtained these relevant observations, a {\em robust} inference requires generating new controls, not present in $\mathcal{O}_q$. This is achieved by creating pairs $\langle x, y\rangle$ similar to those in $\mathcal{O}_q$ but without assuming that they share with the latter a common value of the latent variable. This means that only the {\em observable} features of the entries $o \in \mathcal{O}_q$ must be used to create fictitious controls in $\bar{\mathcal{O}}_q$, for instance as in the examples of Section 5 of \cite{delbianco2021methodology}. \\

\section{Framework}
As said, our procedure takes elements from different approaches:

\begin{itemize}
    
    \item \cite{lei2018distribution} presents three versions of the {\em conformal prediction} method. One is based on the original formulation of  \cite{vovk2005algorithmic}. The other two are the {\em split conformal} and the {\em jackknife} alternatives.
    \item \cite{lei2018distribution} also presents an {\bf R} code for distribution-free prediction, which we modify to be applied in our project.
    \item We use both a classical linear procedure and one based on elastic nets. Alternatively, we can include a Gaussian Kernel regression to implement a non-parametric approach. Other methods can be also easily applied using the estimations and predictions generated by the {\bf R} package \texttt{Caret} (\cite{kuhn2008building}), which is convenient for the following additional reasons: 
    \begin{enumerate}
        \item It implements a previous training stage, cross-validating the results in order to choose the meaningful features. It even allows to apply methodologies that split the entries of the training database in a different way as conformal prediction. This allows to make a finer selection of the $n_r$ relevant entries and the $2n_r$ controls. 
        \item It allows to choose from almost fifty prediction methodologies, among which are GLM, Kernel, Quantile regression, Random Forest, etc. But, as said, we use in a first run only regression and LASSO. 
    \end{enumerate}

 \item The parameters in our exercise are: $\alpha$ for confidence intervals, $\rho$ for splitting in split conformal prediction, and $\gamma$ for cosine similarity (for percentile similarity we also use $\alpha$). 
    \item We add two stages of prediction, using the concept of relevance for individual inference of our original presentation  (\cite{delbianco2021methodology}):
    \begin{enumerate}
        \item The first stage evaluates the relevance of entries in the database for each new observation, based on the degree of similarity with the rest of the database. The relevance is defined, when $p$ is small, in terms of the percentage of distance between the tails of observations and those in the database. Otherwise, when the dimension of the tails is large (in particular when $p>n$), we apply a cosine-based measure. That is, the closer to 1 the cosine between the respective tails is, the more relevant is an entry for a new observation.
        A prediction can be made on the basis of the relevant entries: a possible value of the head of the new observation can be predicted, using a standard prediction method on the subset of relevant entries of the original database.
        \item A second type of prediction is based on the generation of controls using the relevant entries collected in the previous stage. For each individual entry in a query for $X_0$, we generate another tail-head vector by adding a small noise to its features. Thus, given $n_r$ relevant entries ($n_r<<n$), we obtain $2n_r$ control vectors at this stage. 
    \end{enumerate}
\end{itemize}

We will now present the pseudocode of the algorithm that yields the intervals of prediction for the queries. Notice that for simplicity we do not state explicitly the steps and parameters of the methods drawn from $\mathcal{M}$ (\cite{lei2018distribution},  implemented in the {\bf R} package \texttt{ConformalInference}\footnote{\texttt{https://github.com/ryantibs/conformal}}), $\mathcal{A}$ (implemented in the \texttt{Caret} package\footnote{\texttt{https://topepo.github.io/caret/index.html}}) and the similarity functions in $\mathcal{S}$. \\ 



\begin{algorithm}~\label{alg:indcon}
\caption{Individualized Conformal Prediction}
\begin{algorithmic}
\State \textbf{Input}: $X$, $Y$, $X_0$, Regression method $\mathcal{A}$, Similarity method $\mathcal{S}$, Conformal method $\mathcal{M}$, miscoverage level $\alpha$, Similarity intensity $\gamma$
\State \textbf{Output}: Prediction intervals for each element in $X_0$
\State $\alpha \gets \alpha_i \in \{0;1\}$
\State $\gamma \gets \gamma_i \in \{0;1\}$
\State $\mathcal{A} \gets \mathcal{A}_i $ \Comment{OLS, LASSO, Kernel}
\State $\mathcal{M} \gets \mathcal{M}_i$ \Comment{Conformal, Split, Jackkinfe}
\State $\mathcal{S} \gets \mathcal{S}_i$ \Comment{Percentile, Cosine}
\For{\textbf{each} $x \in X_0$}
    \State \textbf{1.1} Predict $y_0$ with $\mathcal{A}$ and $X$
    \State \textbf{1.2} Use residuals and $\mathcal{M}$ $\rightarrow$ interval ($\mathcal{C}$)
    \State \textbf{2.1} Use $\gamma$ and $\mathcal{S}$ $\rightarrow$ $X_{rel}$ 
    \State \textbf{2.2} Predict $y_0$ with $\mathcal{A}$ and $X_{rel}$
    \State \textbf{2.3} Use residuals and $\mathcal{M}$ $\rightarrow$ relevant interval ($\mathcal{C}_{rel}$)
    \State \textbf{3.1} Simulate $X_{sim}$ mimicking the distribution of $X_{rel}$ and $n{sim} = n{rel}$
    \State \textbf{3.2} Predict $y_0$ with $\mathcal{A}$ and $X_{sim}$
    \State \textbf{3.3} Use residuals and $\mathcal{M}$ $\rightarrow$ relevant and simulated interval ($\mathcal{C}_{sim}$)
\EndFor \textbf{ each }$x \in X_0$
\State \textbf{Return: $\mathcal{C}$, $\mathcal{C}_{rel}$, $\mathcal{C}_{sim}$ }

\end{algorithmic}

\end{algorithm}

A diagrammatic representation of this algorithm is depicted in Figure~\ref{fig:diagram}. We can see that there are three paths, leading either to $\mathcal{C}$, $\mathcal{C}_{rel}$ or $\mathcal{C}_{sim}$. Each of those can be executed independently, although the latter two share a stage of detection of relevant observations.\\ 

In Figure~\ref{fig:diagram} the main decisions to be made are represented by the red circle. Each single change in the choices made can lead to different results. In principle we do not know how robust are the results to small differences in the choices made, but some are easily predicted. So, for instance, a lower significance level $\alpha$ yields longer intervals while a higher similarity level $\gamma$ reduces the number of relevant observations. Other than that, the actual choices of estimation method, type of conformal inference and relevance criterion depend on the problems at hand and how the user assesses them.\\

\begin{figure}[hbt!]
		\centering
		\begin{tikzpicture}[->,>=stealth',shorten >=1pt,auto,
	semithick,scale=1]
	\tikzstyle{every state}=[fill=gray!20!white,draw=none]

\tikzset{
    bluenode/.style={
        draw=blue,circle
        },
    rednode/.style={
        circle,draw=purple!80, 
        inner sep=1pt,
        fill=red!20!white,
        },
}
 
	\coordinate (x1) at (0,2);
	\coordinate (x2) at (1.5,5);
	\coordinate (x3) at (8.5,5.5);
	\coordinate (x4) at (11,5.5);
	\coordinate (x5) at (2.5,2);
	\coordinate (x6) at (8.5,3);
	\coordinate (x7) at (11,3);
    \coordinate (x8) at (5.5,0.5);
    \coordinate (x9) at (8.5,0.5);
    \coordinate (x10) at (11,0.5);
 
	\node[state] (N1) at (x1) {\begin{tabular}{c} $D=(X,Y)$ \\ $q= x_0$ \\ \end{tabular}};
	\node[rednode] (N2) at (x2) {\begin{tabular}{c} $\mathcal{A}$ \\ $\mathcal{S}$ \\ $\mathcal{M}$ \\ $\alpha$ \\ $\gamma$ \\ \end{tabular}};
	\node[state] (N3) at (x3) {$\hat{y_0}=\mathcal{A}(X)$};
	\node[state] (N4) at (x4) {$\hat{\mathcal{C}}$};
	\node[state] (N5) at (x5) {$D_{rel}\in D$};
	\node[state] (N6) at (x6) {$\hat{y_0}=\mathcal{A}(X_{rel})$};
	\node[state] (N7) at (x7) {$\hat{\mathcal{C}}_{rel}$};
    \node[state] (N8) at (x8) {$X_{sim} \sim X_{rel}$};
    \node[state] (N9) at (x9) {$\hat{y_0}=\mathcal{A}(X_{sim})$};
	\node[state] (N10) at (x10) {$\hat{\mathcal{C}}_{sim}$};

	\draw[-stealth] (N1) edge (N2) 
	(N2) edge  (N3) 
	(N3) edge (N4)
	(N2) edge (N5)
	(N5) edge (N6)
	(N6) edge (N7)
	(N5) edge (N8) 
	(N8) edge (N9) 
	(N9) edge (N10)
	;
	
\end{tikzpicture}
    \caption{Diagram of Algorithm 1.} 
    \caption*{The red circle encloses the initialization stage of the algorithm in which the estimation method, the type of conformal inference and the relevance criterion are chosen. These decisions impact on all the outputs. Up from this point, three alternative paths can be taken. The first one does not involve relevant data and yields the standard result of conformal prediction, $\mathcal{C}$. The second and the third paths require selecting a subset of observations of $D$ that are similar to the query $x_0$. From then on, two possible alternatives consist in either using just that subset to obtain the confidence interval  $\mathcal{C}_{rel}$ or to generate a synthetic sample enlarging the class of relevant data to obtain $\mathcal{C}_{sim}$. In Algorithm 1 these alternative paths are denoted {\bf 1}, {\bf 2} and {\bf 3}, with their respective steps ({\bf 1.1}, {\bf 1.2}, {\bf 2.1}, etc.)}
    \label{fig:diagram}
\end{figure}

To compare the results of the different methods chosen we apply four metrics. Two of them, $A$ and $B$, do not evaluate the intervals but the predicted values of the variable of interest.

\begin{itemize}

\item[$A$.] {\bf Distance from the forecast}: the absolute value of the difference between the true value $y_0$ and the value predicted by $\mathcal{A}$ on $X$, $X_{rel}$, or $X_{sim}$. Naturally, the smaller $A$ the more accurate the prediction.

\item[$B$.] {\bf Distance from the forecast, as a percentage of $y_0$}: it is defined as $\frac{A}{y_0}$ indicating how far is the forecast from the actual one in proportion to the latter. In this case it is not clear that smaller is always better. 

\item[$C$.] {\bf Length of the interval}: the difference between the upper and lower limits of the intervals generated by Algorithm 1. The length of an interval depends on $\alpha$, but at the same significance level, a shorter interval indicates a lower uncertainty about the forecast.

\item[$D$.] {\bf Normalized distance from the forecast}: it is defined as $\frac{A}{C}$. It indicates how large is the error with respect the length of the interval.

\end{itemize}

Two additional measures may also provide information about the quality of the intervals generated by Algorithm 1, namely {\em coverage} and {\em excess} (see \cite{seedat2023improving}). In the exercises reported in Section 4, all the methods make similar predictions either inside or outside the confidence interval, and thus these metrics may not provide extra information. But for larger heterogeneous databases they may become more informative.\\

\section{Empirical setting}

For our preliminary explorations we generate a dataset of 750 observations, where each third corresponds to a different and heterogeneous data generating process. The corresponding settings are presented in Table~\ref{tab:small}.\\ 

We also enlarge the number of variables in the tails of the entries in the databases ($p$). The description of the models is shown in Table~\ref{tab:long data}.\\

Tables \ref{tab:small cosine raw} and \ref{tab:small percentile raw} show the results of running  Algorithm 1 on the simulated data of Table~\ref{tab:small}. The metrics on these results as well as on those obtained applying Gaussian Kernel and LASSO are presented in Tables~\ref{tab:short_perc}, \ref{tab:short_cos} and \ref{tab:long_cosine}.\\

\begin{figure}[hbt!] 
\begin{subfigure}[h]{0.4\linewidth}
    \centering
    \includegraphics[scale=0.35]{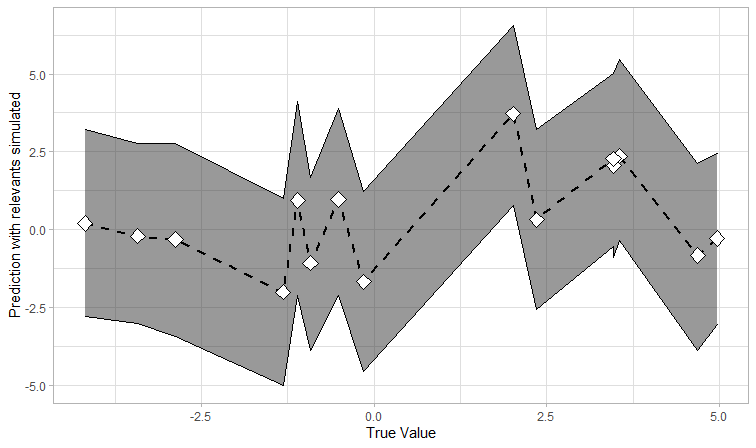}
    \caption{Predictions}
    \label{fig:my_label}
    \end{subfigure}
\hfill
\begin{subfigure}[h]{0.4\linewidth}
    \centering
    \includegraphics[scale=0.35]{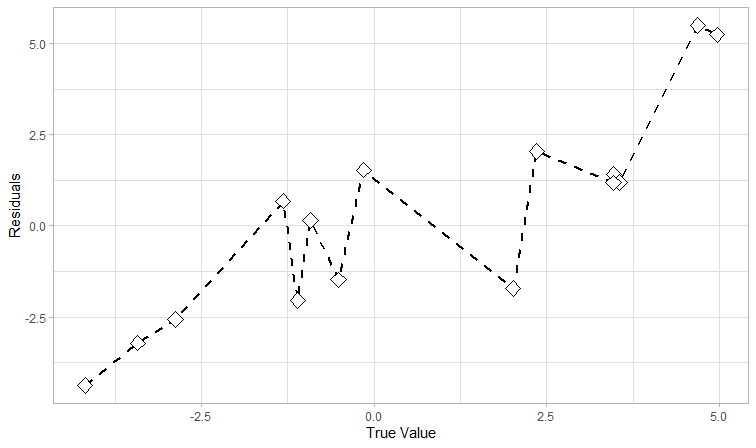}
    \caption{Residuals}
    \label{fig:my_label}
    \end{subfigure}
    \caption{Conformal prediction with simulated relevant controls and OLS}
    \label{fig4}
\end{figure}

\begin{figure}[hbt!]
\begin{subfigure}[h]{0.4\linewidth}
    \centering
    \includegraphics[scale=0.35]{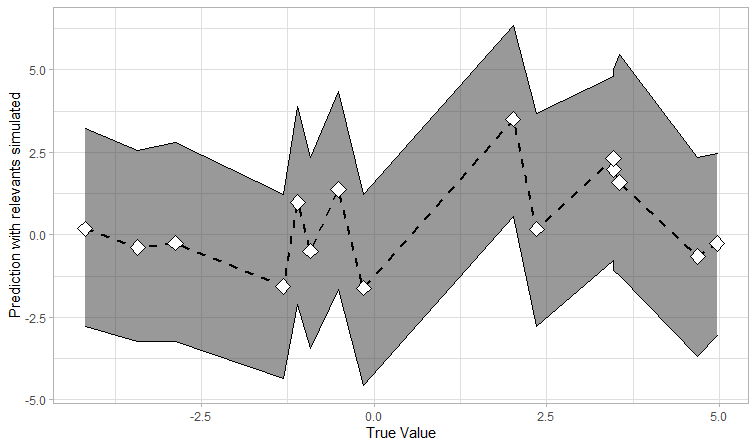}
    \caption{Predictions}
    \label{fig:my_label}
    \end{subfigure}
\hfill
\begin{subfigure}[h]{0.4\linewidth}
    \centering
    \includegraphics[scale=0.35]{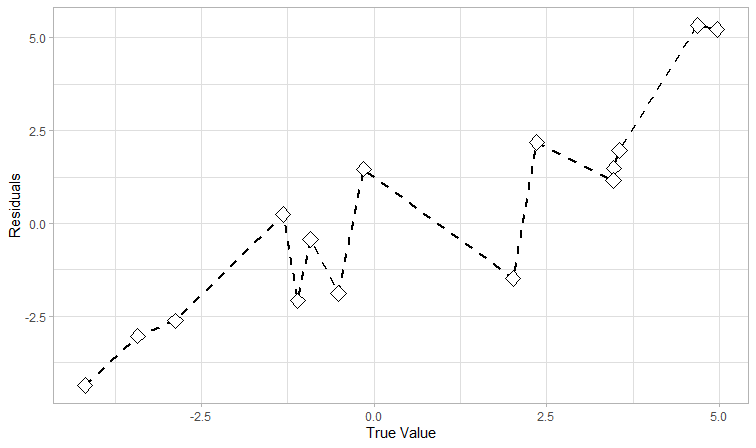}
    \caption{Residuals}
    \label{fig:my_label}
    \end{subfigure}
    \caption{Conformal prediction with simulated relevant controls and LASSO}
    \label{fig5}
\end{figure}

\begin{figure}[hbt!]
\begin{subfigure}[h]{0.4\linewidth}
    \centering
    \includegraphics[scale=0.35]{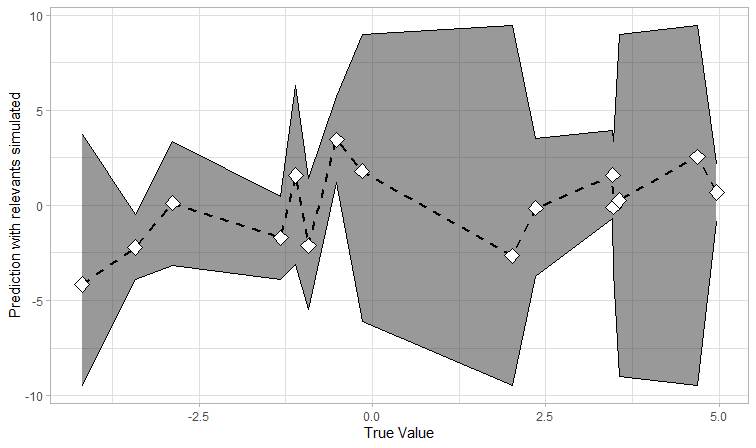}
    \caption{Predictions}
    \label{fig:my_label}
    \end{subfigure}
\hfill
\begin{subfigure}[h]{0.4\linewidth}
    \centering
    \includegraphics[scale=0.35]{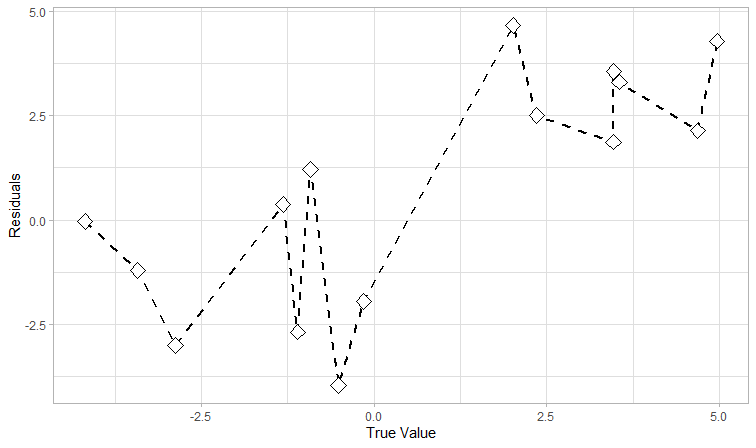}
    \caption{Residuals}
    \label{fig:my_label}
    \end{subfigure}
    \caption{Conformal prediction with simulated relevant controls and Kernel regression}
    \label{fig6}
\end{figure}

Figures~\ref{fig4}, \ref{fig5} and \ref{fig6}  show that the predicted intervals are adapted to the true values. Notice that the residuals increase at the tails of the distribution of true values. This is not quite surprising since we do not use $y_0$ to infer the intervals of the new data used to make the final forecast.\\

\begin{figure}[hbt!]
\begin{subfigure}[h]{0.4\linewidth}
    \centering
    \includegraphics[scale=0.35]{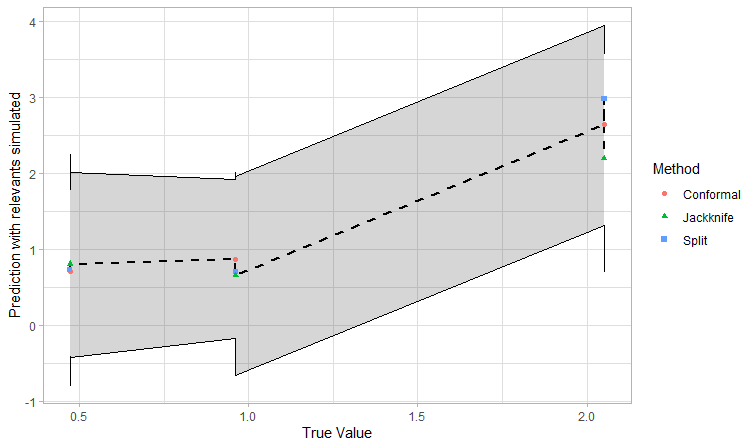}
    \caption{Predictions}
    \label{fig:my_label}
    \end{subfigure}
\hfill
\begin{subfigure}[h]{0.4\linewidth}
    \centering
    \includegraphics[scale=0.35]{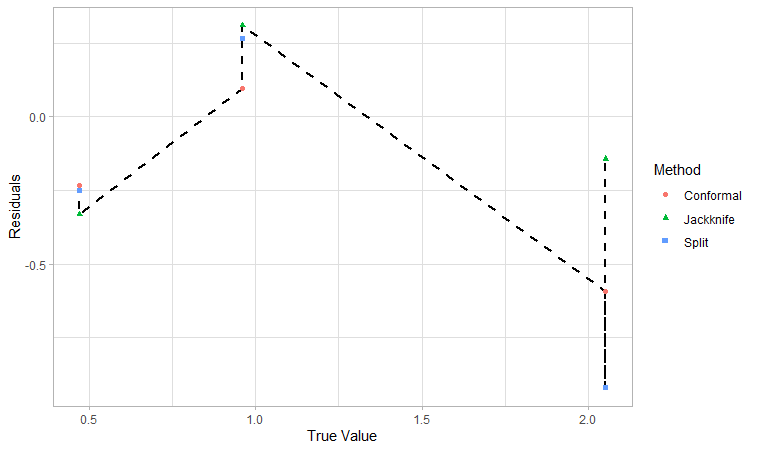}
    \caption{Residuals}
    \label{fig:my_label}
    \end{subfigure}
    \caption{Results of the three methods with simulated relevant controls and LASSO}
    \label{fig7}
\end{figure}

Figure~\ref{fig7} is obtained using the smaller dataset to show the comparison among the three conformal prediction methods. We can see that they do not differ much for the three values used as queries. \\

\section{Discussion}



As already noted, the intervals adapt to each query. The point forecasts depend more on the data used than on the method of conformal prediction chosen. The prediction of the point values tends to be very conservative and thus the residuals become negative on the left and positive on the right. In most cases the predicted intervals capture the true values.\\

The cases analyzed empirically here are, of course, very simple. They do not allow us to explore exhaustively all the consequences of the choices at the initialization phase of Algorithm 1. But we can make the preliminary observation that no clear winner can be ascertained among the conformal prediction method, since the three variants yield in average similar results, although the split conformal one is less computationally costly. \\

Our method yields adaptive intervals using only relevant plus simulated data as to ensure the robustness of the predictions. This {\em divide and conquer} strategy is particularly appropriate for large and heterogeneous datasets. Subsets of data individualize the forecasts of variables of which specific information is not available. \\

As an extra bonus, for each query we obtain a distribution of relevant data. This captures a latent variable (this facilitates the prediction of the intervals corresponding to the query), providing useful information about the individuals in the same ``class''. \\

On the downside, we have seen that residuals are more widely dispersed at the tails of the distribution of actual values. This is inherited from the application of conformal methods, unlike in the self-supervised method of \cite{seedat2023improving}, where residuals are used to train the system to improve its predictions. \\

Further work is needed to make a more precise assessment of the pros and cons of the method, enriching it by relating it to other procedures presented in the literature. Possible lines of study are:

\begin{itemize}
    \item Run more experiments using a larger $p$ or a higher-dimensional setup.
    \item Marginalize the individual result up from that of the the relevant group (or even up from the group of non-relevant ones) as in \cite{zhou2023simulation}.
    \item Explore whether our individualization strategy is compatible with the version of {\em divide and conquer} discussed in \cite{chen2021divide}.
    \item Use residuals to create corrections at the tails of the intervals as to reduce the {\em epistemic uncertainty} at the boundaries \cite{alaa2023conformalized}.
\end{itemize}

\bibliographystyle{apalike}
\bibliography{biblio}

\section*{Appendix}

\begin{table}[hbt!]
    \centering
    \scriptsize

\begin{tabular}{c|c |c}
Setting $A$ & Setting $B$ & Setting $C$ \\ 
one variable  & two variables & heteroskedastic  \\ \hline
$n_1 = 250$  & $n_2 = 250$ & $n_3 = 250$  \\
$u_1 = \mbox{rnorm}(n_1)$  & $u_2 =\mbox{rnorm}(n_2)$ & $x_{1c} = \mbox{rnorm}(n_3,1,1)$  \\
$x_{1a} = \mbox{rnorm}(n_1,1,1)$  & $x_{1b} = \mbox{rnorm}(n_2,3,1)$ & $u_3 = \mbox{rnorm}(n_3)*\frac{x_{1c}}{2}$  \\ 
$x_{2a} = \mbox{rnorm}(n_1,2,1)$ (irrelevant)  & $x_{2b} = \mbox{rnorm}(n_2,2,2)$ & $x_{2c} = \mbox{rnorm}(n_3,3,2)$ \\
$y_a=0.5*x_{1a}+u_1$  & $y_b = 0.5*x_{1b}+0.33*x_{2b}+u_2$ & $y_c = 0.5*x_{1c}*x_{1c}+0.33*x_{2c}+u_3$ \\ \hline
new data for $A$  & new data for $B$ & new data for $C$ \\
$x_{01a} = \mbox{rnorm}(1,1,1)$  & $x_{01b} = \mbox{rnorm}(1,3,1)$ & $x_{01c} = \mbox{rnorm}(1,1,1)$ \\
$x_{02a} = \mbox{rnorm}(1,2,2)$  & $x_{02b} = \mbox{rnorm}(1,2,2)$  & $x_{02c} = \mbox{rnorm}(1,3,2)$ \\
$y_{0a} = 0.5*x_{01a}$   & $y_{0b} =  0.5*x_{01b}+ 0.33*x_{02b}$   & $y_{0c} = 0.5*x_{01c}+0.33*x_{02c}$ \\ 
 $+$ $\mbox{rnorm}(1)$ & $+$ $\mbox{rnorm}(1)$ & $+$ $\mbox{rnorm}(1)* \frac{x_{01c}}{2}$ \\
\hline
    \end{tabular}
    \caption{Simulation datasets with large $n$ and small $p$ (using functions from the {\tt R} language)}
    \label{tab:small}
\end{table}

\begin{table}[!hbt]
    \centering
    \scriptsize
    \begin{tabular}{c |c |c }
$DGP_1$:  & $DGP_2$:  & $DGP_3$:   \\ 
12 variables $N(0,1)$ with 2 coefficients $= 0$ & shift in the mean of the features & shift in the mean of the coefficients   \\ 
$n_1=100; p_1 = 12; s_1 = 2$ & $n_2 = 100; p_2 = 12; s_2 = 2$ & $n_3 = 100; p_3 = 12; s_3 = 2$   \\ 
$x_1 = \mbox{matrix}(\mbox{rnorm}(n_1*p_1),n_1,p_1)$ & $x_2 = \mbox{matrix}(\mbox{rnorm}(n_2*p_2, \mbox{mean}=1),n_2,p_2)$ & $x_3 = \mbox{matrix}(\mbox{rnorm}(n_3*p_3),n_3,p_3)$   \\ 
$\beta_1 = \mbox{c}(\mbox{rnorm}(s_1),\mbox{rep}(0,p_1-s_1))$ & $\beta_2 = \mbox{c}(\mbox{rnorm}(s_2),\mbox{rep}(0,p_2-s_2))$ & $\beta_3 = \mbox{c}(\mbox{rnorm}(s_3, \mbox{mean}=1),\mbox{rep}(0,p_3-s_3))$   \\ 
$y_1 = x_1 * \beta_1 + \mbox{rnorm}(n_1)$ & $y_2 = x_2 * \beta_2 + \mbox{rnorm}(n_2)$ & $y_3 = x_3 * \beta_3 + \mbox{rnorm}(n_3)$   \\ 
$n_{01} = 5$ & $n_{02} = 5$ & $n_{03} = 5$  \\ 
$x_{01} = \mbox{matrix}(\mbox{rnorm}(n_{01}*p_1),n_{01},p_1)$ & $x_{02} = \mbox{matrix}(\mbox{rnorm}(n_{02}*p_2,\mbox{mean}=1),n_{02},p_2)$ & $x_{03} = \mbox{matrix}(\mbox{rnorm}(n_{03}*p_3),n_{03},p_3)$   \\ 
$y_{01} = x_{01} * \beta_1 + \mbox{rnorm}(n_{01})$ & $y_{02} = x_{02} * \beta_2 + \mbox{rnorm}(n_{02})$ & $y_{03} = x_{03} * \beta_3 + \mbox{rnorm}(n_{03})$   \\ 
$D_1 = \mbox{as.data.frame}(\mbox{cbind}(y_1,x_1))$ & $D_2 = \mbox{as.data.frame}(\mbox{cbind}(y_2,x_2))$ & $D_3 = \mbox{as.data.frame}(\mbox{cbind}(y_3,x_3))$   \\ 
$x_1 = \mbox{as.data.frame}(x_1)$ & $x_2 = \mbox{as.data.frame}(x_2)$ & $x_3 = \mbox{as.data.frame}(x_3)$   \\ 

    \end{tabular}
    \caption{Simulation datasets with smaller $n$ and large $p$ (using functions from the {\tt R} language)}
    \label{tab:long data}
\end{table}

\begin{table}[hbt!]
    \centering
    \begin{tabular}{|l|c c c|c c c|c c c |}
    \hline
Cosine & \multicolumn{3}{|c|}{Confomal}  & \multicolumn{3}{|c|}{Split}  & \multicolumn{3}{|c|}{Jackknife} \\ \hline
variable & $1$ & $2$ & $3$ & $1$ & $2$ & $3$ & $1$ & $2$ & $3$\\ \hline
y0 & $0.96$ & $2.05$ & $0.47$ & $0.96$ & $2.05$ & $0.47$ & $0.96$ & $2.05$ & $0.47$\\ \hline
pred & $0.96$ & $2.59$ & $0.95$ & $0.96$ & $2.79$ & $0.9$ & $2.06$ & $0.71$ & $0.98$\\
predr & $0.88$ & $2.43$ & $0.84$ & $0.79$ & $2.81$ & $0.75$ & $0.67$ & $2.19$ & $0.84$\\
predrs & $0.88$ & $2.41$ & $0.84$ & $0.91$ & $2.63$ & $0.9$ & $0.69$ & $2.2$ & $0.8$\\
predl & $0.96$ & $2.59$ & $0.95$ & $0.96$ & $2.59$ & $0.95$ & $2.05$ & $0.71$ & $0.99$\\
predlr & $0.88$ & $2.43$ & $0.84$ & $0.79$ & $2.81$ & $0.75$ & $0.64$ & $2.08$ & $0.78$\\
predlrs & $0.9$ & $2.47$ & $0.86$ & $0.91$ & $2.63$ & $0.9$ & $0.65$ & $2.19$ & $0.8$\\ \hline
lo & $-0.36$ & $1.36$ & $-0.36$ & $-0.83$ & $1.01$ & $-0.89$ & $0.22$ & $-1.13$ & $-0.85$\\
lor & $0.65$ & $0.79$ & $1.08$ & $-1.28$ & $0.36$ & $-0.88$ & $-0.73$ & $0.59$ & $-0.48$\\
lors & $-0.36$ & $1.08$ & $-0.36$ & $-0.59$ & $0.49$ & $-0.08$ & $-0.63$ & $0.71$ & $-0.42$\\
lol & $-0.36$ & $1.36$ & $-0.36$ & $-0.81$ & $0.83$ & $-0.81$ & $0.22$ & $-1.12$ & $-0.84$\\
lolr & $0.5$ & $0.79$ & $1.08$ & $-1.28$ & $0.37$ & $-0.88$ & $-0.73$ & $-0.41$ & $-0.62$\\
lolrs & $-0.36$ & $1.08$ & $-0.36$ & $-0.59$ & $0.49$ & $-0.08$ & $-0.67$ & $0.71$ & $-0.42$\\ \hline
up & $2.22$ & $3.8$ & $2.22$ & $2.74$ & $4.58$ & $2.69$ & $3.89$ & $2.54$ & $2.82$\\
upr & $3.08$ & $3.51$ & $3.51$ & $2.96$ & $4.52$ & $2.39$ & $2.08$ & $3.79$ & $2.15$\\
uprs & $2.08$ & $3.8$ & $2.08$ & $2.39$ & $4.51$ & $1.88$ & $2$ & $3.69$ & $2.03$\\
upl & $2.22$ & $3.8$ & $2.22$ & $2.72$ & $4.36$ & $2.71$ & $3.88$ & $2.54$ & $2.82$\\
uplr & $3.08$ & $3.51$ & $3.51$ & $2.96$ & $4.53$ & $2.39$ & $2$ & $4.57$ & $2.19$\\
uplrs & $2.22$ & $3.8$ & $2.08$ & $2.39$ & $4.51$ & $1.88$ & $1.95$ & $3.57$ & $2.01$\\ \hline
    \end{tabular}
    \caption{Cosine Relevance}
    \caption*{\textbf{Note}: \textit{pred}, \textit{lo} and \textit{up} denote the prediction, lower limit and upper limit of the interval, respectively. In turn, \textit{r} correspond to the case when only relevant observations are used while \textit{rs} appears in cases in which both relevant and simulated observations are included. Finally, \textit{l} and \textit{k} refer to the cases in which LASSO regressions and kernel estimations are used, respectively.}
    \label{tab:small cosine raw}
\end{table}

\begin{table}[hbt!]
    \centering
    \begin{tabular}{|l|c c c|c c c|c c c |}
    \hline
 Percentile & \multicolumn{3}{|c|}{Confomal}  & \multicolumn{3}{|c|}{Split}  & \multicolumn{3}{|c|}{Jackknife} \\ \hline
variable & $1$ & $2$ & $3$ & $1$ & $2$ & $3$ & $1$ & $2$ & $3$\\ \hline
y0 & $0.96$ & $2.05$ & $0.47$ & $0.96$ & $2.05$ & $0.47$ & $0.96$ & $2.05$ & $0.47$\\ \hline
pred & $0.96$ & $2.59$ & $0.95$ & $0.96$ & $2.79$ & $0.9$ & $2.06$ & $0.71$ & $0.98$\\ 
predr & $0.84$ & $2.57$ & $0.69$ & $0.43$ & $2.05$ & $0.58$ & $0.67$ & $2.19$ & $0.84$\\
predrs & $0.86$ & $2.56$ & $0.68$ & $0.7$ & $2.97$ & $0.72$ & $0.69$ & $2.2$ & $0.8$\\
predl & $0.96$ & $2.59$ & $0.95$ & $0.96$ & $2.59$ & $0.95$ & $2.05$ & $0.71$ & $0.99$\\
predlr & $0.84$ & $2.57$ & $0.69$ & $0.43$ & $2.02$ & $0.58$ & $0.64$ & $2.08$ & $0.78$\\
predlrs & $0.87$ & $2.64$ & $0.7$ & $0.7$ & $2.97$ & $0.72$ & $0.65$ & $2.19$ & $0.8$\\ \hline
lo & $-0.36$ & $1.36$ & $-0.36$ & $-0.34$ & $1.5$ & $-0.39$ & $0.22$ & $-1.13$ & $-0.85$\\
lor & $0.76$ & $0.97$ & $-0.28$ & $-0.21$ & $1.25$ & $-0.59$ & $-0.73$ & $0.59$ & $-0.48$\\
lors & $-0.25$ & $1.2$ & $-0.41$ & $-0.34$ & $0.88$ & $-0.79$ & $-0.63$ & $0.71$ & $-0.42$\\
lol & $-0.36$ & $1.36$ & $-0.36$ & $-0.31$ & $1.33$ & $-0.32$ & $0.22$ & $-1.12$ & $-0.84$\\
lolr & $1.05$ & $1.2$ & $-0.47$ & $-0.26$ & $1.25$ & $-0.6$ & $-0.73$ & $-0.41$ & $-0.62$\\
lolrs & $-0.18$ & $1.31$ & $-0.41$ & $-0.34$ & $0.88$ & $-0.79$ & $-0.67$ & $0.71$ & $-0.42$\\ \hline
up & $2.22$ & $3.8$ & $2.22$ & $2.25$ & $4.09$ & $2.19$ & $3.89$ & $2.54$ & $2.82$\\
upr & $2.92$ & $3.49$ & $1.91$ & $2.1$ & $4.09$ & $1.75$ & $2.08$ & $3.79$ & $2.15$\\
uprs & $1.91$ & $3.94$ & $1.79$ & $2$ & $3.9$ & $2.24$ & $2$ & $3.69$ & $2.03$\\
upl & $2.22$ & $3.8$ & $2.22$ & $2.22$ & $3.86$ & $2.21$ & $3.88$ & $2.54$ & $2.82$\\
uplr & $3.29$ & $3.83$ & $1.72$ & $2.17$ & $4.09$ & $1.76$ & $2$ & $4.57$ & $2.19$\\
uplrs & $1.91$ & $3.94$ & $1.79$ & $2$ & $3.9$ & $2.24$ & $1.95$ & $3.57$ & $2.01$\\ \hline
    \end{tabular}
    \caption{Percentile relevance}
    \caption*{\textbf{Note}: Here we use the same abbreviations as in table \ref{tab:small cosine raw}}
    \label{tab:small percentile raw}
\end{table}

\begin{table}[!hbt]
    \centering
    \begin{tabular}{l|c|c|c|c}
     \multicolumn{5}{c}{Short data}\\
Percentile & General & Conformal & Split & Jackknife \\ \hline
diffpred & 0.57 & 0.34 & 0.39 & 0.98 \\
diffpredr & 0.25 & 0.28 & 0.21 & 0.26 \\
diffpredrs & 0.33 & 0.27 & 0.48 & 0.25 \\
diffpredl & 0.55 & 0.34 & 0.34 & 0.98 \\
diffpredlr & 0.24 & 0.28 & 0.22 & 0.22 \\
diffpredlrs & 0.35 & 0.31 & 0.48 & 0.26 \\
diffpredk & 0.32 & 0.3 & 0.31 & 0.34 \\
diffpredkr & 0.24 & 0.2 & 0.24 & 0.28 \\
diffpredkrs & 0.25 & 0.19 & 0.27 & 0.28 \\ \hline 
\%pred & 60.25 & 42.49 & 42.38 & 95.87 \\
\%predr & 30.52 & 27.57 & 26.09 & 37.92 \\
\%predrs & 34.42 & 26.03 & 41.81 & 35.41 \\
\%predl & 60.31 & 42.49 & 42.56 & 95.89 \\
\%predlr & 29.28 & 27.57 & 26.65 & 33.62 \\
\%predlrs & 35.9 & 29.26 & 41.91 & 36.54 \\
\%predk & 38.37 & 36.05 & 36.49 & 42.58 \\
\%predkr & 26.22 & 23.44 & 27.83 & 27.39 \\
\%predkrs & 27.79 & 23.29 & 32.62 & 27.45 \\ \hline
int & 2.93 & 2.53 & 2.59 & 3.67 \\
intr & 2.56 & 2.29 & 2.49 & 2.89 \\
intrs & 2.62 & 2.37 & 2.8 & 2.69 \\
intl & 2.91 & 2.53 & 2.53 & 3.66 \\
intlr & 2.8 & 2.35 & 2.54 & 3.5 \\
intlrs & 2.58 & 2.31 & 2.8 & 2.64 \\
intk & 2.56 & 2.53 & 2.49 & 2.65 \\
intkr & 2.95 & 2.56 & 2.61 & 3.66 \\
intkrs & 2.45 & 2.29 & 2.65 & 2.41 \\ \hline
ab & 0.6 & 0.63 & 0.65 & 0.52 \\
abr & 0.61 & 0.71 & 0.59 & 0.53 \\
abrs & 0.55 & 0.58 & 0.55 & 0.53 \\
abl & 0.6 & 0.63 & 0.63 & 0.52 \\
ablr & 0.62 & 0.76 & 0.59 & 0.5 \\
ablrs & 0.55 & 0.59 & 0.55 & 0.51 \\
abk & 0.62 & 0.61 & 0.62 & 0.63 \\
abkr & 0.57 & 0.63 & 0.54 & 0.55 \\
abkrs & 0.54 & 0.54 & 0.51 & 0.58 \\ \hline  
    \end{tabular}
    \caption{Short data percentile}
    \caption*{\textbf{Note}: \textit{diffpred} denotes the differences between the predicted and the true values. \textit{\%pred} indicates the same difference but relative to the true value. \textit{int} is the length of the interval (the upper minus the lower limits). Finally, \textit{ab} represents the size of the error relative to the length of the interval.}
    \label{tab:short_perc}
\end{table}

\begin{table}[!hbt]
    \centering
    \begin{tabular}{l|c|c|c|c}
\multicolumn{5}{c}{Short data}\\
Cosine & General & Conformal & Split & Jackknife \\ \hline
diffpred & 0.57 & 0.34 & 0.39 & 0.98 \\
diffpredr & 0.31 & 0.28 & 0.4 & 0.26 \\
diffpredrs & 0.29 & 0.27 & 0.35 & 0.25 \\
diffpredl & 0.55 & 0.34 & 0.34 & 0.98 \\
diffpredlr & 0.3 & 0.28 & 0.4 & 0.22 \\
diffpredlrs & 0.3 & 0.29 & 0.35 & 0.26 \\
diffpredk & 0.32 & 0.3 & 0.34 & 0.31 \\
diffpredkr & 0.6 & 0.8 & 0.57 & 0.44 \\
diffpredkrs & 0.93 & 1.1 & 0.58 & 1.11 \\ \hline
\%pred & 60.25 & 42.49 & 42.38 & 95.87 \\
\%predr & 37.08 & 35.18 & 38.15 & 37.92 \\
\%predrs & 37.27 & 34.79 & 41.6 & 35.41 \\
\%predl & 60.31 & 42.49 & 42.56 & 95.89 \\
\%predlr & 35.65 & 35.18 & 38.15 & 33.62 \\
\%predlrs & 38.18 & 36.4 & 41.6 & 36.54 \\
\%predk & 38.41 & 36.05 & 42.7 & 36.49 \\
\%predkr & 89.13 & 102.14 & 99.06 & 66.2 \\
\%predkrs & 144.67 & 153.04 & 99.05 & 181.92 \\ \hline
int & 3.26 & 2.53 & 3.58 & 3.67 \\
intr & 3.1 & 2.53 & 3.89 & 2.89 \\
intrs & 2.74 & 2.53 & 2.99 & 2.69 \\
intl & 3.24 & 2.53 & 3.52 & 3.66 \\
intlr & 3.33 & 2.58 & 3.89 & 3.5 \\
intlrs & 2.74 & 2.58 & 2.99 & 2.64 \\
intk & 2.56 & 2.53 & 2.64 & 2.49 \\
intkr & 2.93 & 2.86 & 3.41 & 2.52 \\
intkrs & 2.91 & 3.37 & 2.68 & 2.67 \\ \hline
ab & 0.59 & 0.63 & 0.61 & 0.52 \\
abr & 0.65 & 0.88 & 0.55 & 0.53 \\
abrs & 0.57 & 0.59 & 0.6 & 0.53 \\
abl & 0.58 & 0.63 & 0.6 & 0.52 \\
ablr & 0.64 & 0.87 & 0.55 & 0.5 \\
ablrs & 0.57 & 0.6 & 0.6 & 0.51 \\
abk & 0.62 & 0.61 & 0.63 & 0.62 \\
abkr & 0.62 & 0.37 & 0.54 & 0.94 \\
abkrs & 0.73 & 0.62 & 0.58 & 0.99 \\ \hline
       
    \end{tabular}
    \caption{Short data cosine}
    \caption*{\textbf{Note}: Here we use the same abbreviations as in table \ref{tab:short_perc}}
    \label{tab:short_cos}
\end{table}

\begin{table}[!hbt]
    \centering
    \begin{tabular}{l|c|c|c|c}
 \multicolumn{5}{c}{Long data}\\ 
Cosine & General & Conformal & Split & Jackknife \\ \hline
diffpred & 2.8 & 2.47 & 2.58 & 3.34 \\
diffpredr & 2.75 & 2.27 & 2.64 & 3.34 \\
diffpredrs & 2.8 & 2.29 & 2.76 & 3.34 \\
diffpredl & 2.81 & 2.48 & 2.6 & 3.35 \\
diffpredlr & 2.73 & 2.29 & 2.61 & 3.28 \\
diffpredlrs & 2.78 & 2.34 & 2.69 & 3.33 \\
diffpredk & 2.05 & 1.5 & 2.32 & 2.33 \\
diffpredkr & 1.91 & 1.7 & 1.87 & 2.17 \\
diffpredkrs & 2.1 & 1.77 & 2.39 & 2.14 \\ \hline
\%pred & 179.03 & 165.13 & 105.96 & 266.02 \\
\%predr & 168.86 & 157.93 & 115.92 & 232.74 \\
\%predrs & 183.81 & 157.16 & 165.13 & 229.12 \\
\%predl & 176.34 & 165.17 & 98.11 & 265.75 \\
\%predlr & 173.31 & 135.75 & 109.4 & 274.78 \\
\%predlrs & 174.9 & 160.28 & 136.48 & 227.92 \\
\%predk & 161.98 & 132.38 & 183.64 & 169.93 \\
\%predkr & 131.49 & 166.7 & 109.41 & 118.35 \\
\%predkrs & 154.95 & 126.67 & 215.02 & 123.15 \\ \hline
int & 5.13 & 5.86 & 5.96 & 3.57 \\
intr & 5.03 & 5.51 & 5.95 & 3.64 \\
intrs & 4.31 & 5.87 & 3.52 & 3.54 \\
intl & 5.14 & 5.69 & 6.38 & 3.36 \\
intlr & 6.78 & 5.76 & 6.24 & 8.33 \\
intlrs & 4.31 & 5.94 & 3.52 & 3.48 \\
intk & 5.43 & 4.12 & 6.71 & 5.45 \\
intkr & 6.05 & 5.27 & 5.58 & 7.31 \\
intkrs & 5.97 & 1.77 & 9.41 & 6.73 \\ \hline
ab & 0.6 & 0.42 & 0.43 & 0.94 \\
abr & 0.59 & 0.41 & 0.45 & 0.92 \\
abrs & 0.71 & 0.39 & 0.8 & 0.94 \\
abl & 0.61 & 0.44 & 0.41 & 1 \\
ablr & 0.41 & 0.4 & 0.42 & 0.4 \\
ablrs & 0.75 & 0.39 & 0.78 & 1.09 \\
abk & 0.38 & 0.36 & 0.35 & 0.43 \\
abkr & 0.33 & 0.33 & 0.35 & 0.3 \\
abkrs & 0.63 & 1.21 & 0.36 & 0.32 \\ \hline     
    \end{tabular}
    \caption{Long data cosine}
    \caption*{\textbf{Note}: Here we use the same abbreviations as in table \ref{tab:short_perc}}
    \label{tab:long_cosine}
\end{table}

\end{document}